\documentclass[english,aps,superscriptaddress,preprintnumbers,reprint,footinbib,amsmath,amssymb,prb]{revtex4-2}
\usepackage[latin9]{inputenc}
\usepackage{babel}
\usepackage{amsmath}
\usepackage{amssymb}
\usepackage{graphicx}
\usepackage{esint}
\usepackage{hyperref}

\makeatletter

\@ifundefined{textcolor}{}{%
 \definecolor{BLACK}{gray}{0}
 \definecolor{WHITE}{gray}{1}
 \definecolor{RED}{rgb}{1,0,0}
 \definecolor{GREEN}{rgb}{0,1,0}
 \definecolor{BLUE}{rgb}{0,0,1}
 \definecolor{CYAN}{cmyk}{1,0,0,0}
 \definecolor{MAGENTA}{cmyk}{0,1,0,0}
 \definecolor{YELLOW}{cmyk}{0,0,1,0}
}

\usepackage{aecompl}

\usepackage{epsfig}\usepackage{dcolumn}\usepackage{bm}
\usepackage{babel}

\hypersetup{
    colorlinks=true,
    linkcolor=blue,
    urlcolor=blue,
    citecolor=blue
        }

\makeatother

\begin{document}
\title{Spin-Exchange Induced Spillover on Poor Man's Majoranas in Minimal
Kitaev Chains}
\author{J.E. Sanches}
\email[corresponding author:]{jose.sanches@unesp.br}

\affiliation{S\~ao Paulo State University (Unesp), School of Engineering, Department
of Physics and Chemistry, 15385-007, Ilha Solteira-SP, Brazil}
\author{L.T. Lustosa}
\affiliation{S\~ao Paulo State University (Unesp), School of Engineering, Department
of Physics and Chemistry, 15385-007, Ilha Solteira-SP, Brazil}
\author{L.S. Ricco}
\affiliation{Science Institute, University of Iceland, Dunhagi-3, IS-107, Reykjavik,
Iceland}
\author{H. Sigur\dh sson}
\affiliation{Science Institute, University of Iceland, Dunhagi-3, IS-107, Reykjavik,
Iceland}
\affiliation{Institute of Experimental Physics, Faculty of Physics, University
of Warsaw, ulica Pasteura 5, PL-02-093 Warsaw, Poland}
\author{\\M. de Souza}
\affiliation{S\~ao Paulo State University (Unesp), IGCE, Department of Physics, 13506-970,
Rio Claro-SP, Brazil}
\author{M.S. Figueira}
\affiliation{Instituto de F\'isica, Universidade Federal Fluminense, 24210-340, Niter\'oi,
Rio de Janeiro, Brazil}
\author{E. Marinho Jr.}
\email[corresponding author:]{enesio.marinho@unesp.br}

\affiliation{S\~ao Paulo State University (Unesp), School of Engineering, Department
of Physics and Chemistry, 15385-007, Ilha Solteira-SP, Brazil}
\author{A.C. Seridonio}
\email[corresponding author:]{antonio.seridonio@unesp.br}

\affiliation{S\~ao Paulo State University (Unesp), School of Engineering, Department
of Physics and Chemistry, 15385-007, Ilha Solteira-SP, Brazil}
\begin{abstract}
The ``Poor Man's Majoranas'' (PMMs) {[}\href{https://journals.aps.org/prb/abstract/10.1103/PhysRevB.86.134528}{Phys. Rev. B 86, 134528 (2012)}{]}
devoid of topological protection can ``spill over'' from one edge
into another of the minimal Kitaev chain when perturbed electrostatically.
As aftermath, this leads to a delocalized Majorana fermion (MF) at
both the edges. Additionally, according to recent differential conductance
measurements in a pair of superconducting and spinless quantum dots
(QDs), such a PMM picture was brought to reality {[}\href{https://www.nature.com/articles/s41586-022-05585-1}{Nature 614, 445 (2023)}
and \href{https://www.nature.com/articles/s41586-024-07434-9}{Nature 630, 329 (2024)}{]}.
Based on this scenario, we propose the spillover of the PMM when its
QD is exchange coupled to a quantum spin $S$. We show that if this
QD is perturbed by the exchange coupling $J$, solely the half $2S+1$
$(2S+2)$ of the fine structure stays explicit for a fermionic (bosonic)
$S.$ Concurrently, the other half squeezes itself as the delocalized
MF zero-mode. Particularly, turning-off the superconductivity the
multiplicity $2S+1$ holds regardless the spin statistics. Meanwhile,
the PMM spillover induced by $J$ becomes a statistics dependent effect.
Hence, our findings contribute to the comprehension of spin-phenomena interplay
with superconductivity in minimal Kitaev chains, offering insights
for future quantum computing devices hosting PMMs.
\end{abstract}
\maketitle

\section{Introduction}

Predicted by Ettore Majorana in 1937, from a particular solution of
the Dirac equation, Majorana Fermions (MFs) are particles identical
to their antiparticles~\cite{Majorana-1937}. These particles were
never observed in experiments. However, in condensed matter, they
can manifest as quasiparticle excitations called Majorana bound states
(MBSs), which appear as zero-energy modes at the edges of topological
superconductors~\cite{Marra-2022,Leakage-2014,Yuval-Oreg-2022,Oppen-2014,Oppen-2010,Oppen-2013,Klinovaja-2013,Yazdani-2013,Fu-Kane-2009,Fu-Kane-2008,M.-Franz-2013,Nagaosa-2013,Pascal-Simon-2013,Potter-Lee-2012,S-CZhang-2011,Beenakker-2015,Fujimoto-2016,YoichiAndo-2017,Yuval-Oreg-2019,Alicea-2012,Beenakker-2013,Flensberg-2021,Klinovaja-2021,C.Marcus-2016,Mourik-2012,C.Marcus-2017}.
{Theoretically, the appearance of these MBSs can be understood through
the Kitaev toy model~\cite{Kitaev-2001}, which describes a one-dimensional
spinless chain with a \textit{p}-wave superconducting pairing symmetry.
In this model, MBSs emerge as topologically protected,
non-local states located at opposite ends of the chain. In the experimental
scenario, the achievement of such quasiparticles is still challenging. Thus, MBSs
are expected to be the building blocks for the highly-pursued fault-tolerant
quantum computing~\cite{Aguado-2017,Flensberg-2021,Marra-2022}.}
Consequently, the last decade has witnessed a plethora of theoretical
and experimental efforts to explore potential MBS hosts as a basis
for cutting-edge quantum computing systems~\cite{Kitaev-2001,Alicea-2012,Colloquium-Franz-2015,Aguado-2017,Yuval-Oreg-2022,Flensberg-2012,Tewari-2013,Flensberg-2021,Yazdani-2021,Marra-2022}.

Realizations of the Kitaev chain have been theoretically proposed
and experimentally implemented in hybrid setups that combine Zeeman
fields, spin-orbit interaction (SOI), and \textit{s}-wave superconducting
proximity effects~\cite{Marra-2022,Aguado-2017}. Examples include
semiconducting nanowires with strong Rashba SOI, such as GaAs and
InSb, proximitized by conventional \textit{s}-wave superconductors~\cite{C.Marcus-2016,Mourik-2012,C.Marcus-2017,Alicea-2012,Colloquium-Franz-2015,Aguado-2017,DasSarma-2010,TopoGapProtocolPhysRevB.107.245423(2023)},
as well as one-dimensional magnetic chains deposited also on conventional
superconductors~\cite{Klinovaja-2013,Pawlak_2016,J.Franke-2015,Beenakker-2011,Yazdani-2013,Pascal-Simon-2013,Yazdani-2014,Yazdani-2017,Nitta-2019,Ernst-Meyer-2019,Yazdani-2021,swaveCoupling,Roland1,Roland2},
all serving as experimental platforms for hosting MBSs.

Despite clear theoretical frameworks for realizing MBSs in the laboratory,
their detection remains challenging, as other phenomena can mimic
MBS tunneling transport signatures. Factors such as disorder and zero-energy-pinned
Andreev reflections~\cite{Flensberg-2021,Marra-2022} can, for instance,
replicate the zero-energy behavior of topological MBSs in a topologically
trivial regime, hampering an unambiguous identification. {Noteworthy,
researchers from Microsoft have claimed recently the realization of
an architecture for a topological quantum computer, but with interferometric
measurements unable to fully distinguish topological from trivial
end states~\cite{Majorana1Nature2025}.}

{As aftermath of several accomplishment issues to
manufacture a long Kitaev topological wire, we focus on a simpler
and experimentally more accessible alternative to hybrid nanowires
and magnetic atom chains, such as linear arrays of quantum dots (QDs) with
superconductivity, which have emerged as a promising approach to realize
shorter Kitaev chains~\cite{Kouwenhoven2023,MinimalModel2,MinimalModel3}.}
This concept was first proposed in seminal works by Jay D. Sau and
S. Das Sarma~\cite{sau2012realizing} and by M. Leijnse and K. Flensberg~\cite{Flensberg_2012(Poor)},
with the latter authors coining the term ``Poor Man's Majoranas''
(PMMs) for MBSs in these QD chains, due to their lack of topological
protection~\cite{Flensberg_2012(Poor),Poor2,Poor3,Poor4,Jelena1,Jelena2,alvarado2024interplaymajoranashibastates}.
Recent experimental realizations on two~\cite{Kouwenhoven2023,MinimalModel2}
and three QD arrays~\cite{MinimalModel3} have revealed electronic
transport signatures consistent with the presence of PMMs.

\begin{figure}[!]
\centering\includegraphics[width=1\columnwidth]{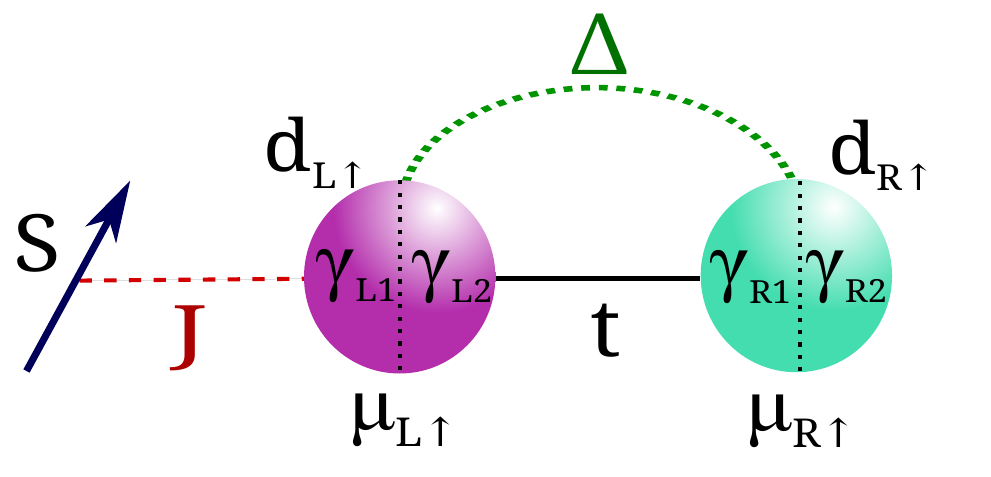} \caption{\label{Fig:Fig.1} {Sketch of the minimal Kitaev chain
with a quantum spin $S$ exchange coupled via $J$ (red dashed line) to the left quantum
dot (purple sphere) with chemical potential $\mu_{L\uparrow},$ fermionic
operator $d_{L\uparrow},$ and the Majorana bound states $\gamma_{L1}$
and $\gamma_{L2}.$ For the right quantum dot (green sphere), we have
similarly $\mu_{R\uparrow},$ $d_{R\uparrow},$ $\gamma_{R1}$ and
$\gamma_{R2}.$ Both the quantum dots are connected to each other
by electron co-tunneling and crossed Andreev reflection given by $t$ (black solid line)
and $\Delta$ (green dotted line) terms, respectively. For $\Delta=t,\mu_{L\uparrow}=\mu_{R\uparrow}=0$
and $J\protect\neq0,$ the spillover of the \textquotedblleft Poor
Man's Majorana\textquotedblright}\textit{{{} }}{$\gamma_{L1}$
yields a delocalized Majorana fermion zero-mode from the squeezing
of the half $2S+1$ $(2S+2)$ of the fine structure due to a fermionic
(bosonic) $S.$ Without crossed Andreev reflection, we have the multiplicity
$2S+1$ statistics independent. With crossed Andreev reflection, the
dependence on $S$ allows to distinguish the spin statistics.}}
\end{figure}

Particularly, in a system of two spin-polarized semiconductor QDs
coupled via an \textit{s}-wave superconductor, known as the minimal
Kitaev chain~\cite{Kouwenhoven2023}, PMMs can form when the QDs
are grounded and symmetrically linked through electronic co-tunneling
(ECT) and crossed Andreev reflection (CAR) processes. At this \textit{sweet
spot}\cite{Flensberg_2012(Poor)}, the PMMs emerge as non-local states,
spatially separated across the two QDs, similar to conventional MBSs.
However, without topological protection, a perturbation to one QD,
such as an applied electrostatic potential, induces the PMM at that
dot to ``spill over'' into the other QD~\cite{Flensberg_2012(Poor)},
resulting in Majorana quasiparticle components in both dots.

{Interestingly enough, the absence of topological
protection of the PMMs is indeed a trump, favoring the initialization
and readout of a PMM-based qubit state~\cite{Poor3}. In opposing, perfect MBSs require the break down of the topological protection
to be manipulated as a qubit state~\cite{BreakingTopo1,BreakingTopo2,BreakingTopo3}. Thus, due to the relative easy way of state preparation
and access via PMMs, braiding and fusion protocols that employ two PMM
setups as }\textit{{Majorana}}{{} qubits
(four }\textit{{Majoranas}}{{} to compose
a two-level system) in a similar architecture to the Tetron
qubit from Microsoft~\cite{RoadmapTetron2025}, have been proposed to verify non-Abelian
Physics~\cite{Poor3}. This is allowed by the qubit representation, which is based
on the fermion parity~\cite{Flensberg_2012(Poor),FermionParity}, i.e., the vacuum (even) or single occupied
(odd) states from the ordinary fermion made by a pair of PMMs, whose
characterization strongly depends on the }\textit{{Majorana}}{{}
quality (}\textit{{Majorana}}{{} polarization),
a measure of the degree of non-locality of these PMMs over the host
QDs~\cite{Poor4}. A detailed discussion on these concepts and near-future experiments covering
PMMs can be found in Ref.~\cite{Poor3}.}

In this work, we benefit from the PMMs lack of topological protection
to shown that the minimal Kitaev chain is useful to distinguish between
fermionic and bosonic spin statistics. We reveal that the PMM exhibits
the spillover characteristic if its QD is exchange coupled to a quantum
spin. By perturbing this QD via variations of the exchange coupling,
the delocalized MF zero-mode arising from the PMM spillover stays
decoupled from the half of the fine structure, which is given by $2S+1$
$(2S+2)$ due to a fermionic (bosonic) spin $S.$ Therefore, the other
half then squeezes the mode at zero energy and ensures its pinning
there. As the fine structure changes with the statistics of $S$ in
the presence of CAR, the PMM spillover by the exchange coupling is
then revealed as a statistics dependent phenomenon. Our findings shed
light on the spillover mechanism of PMMs based on a quantum spin,
thus opening novel possibilities for spin-related and quantum computing
devices.

\section{The Model}

We consider the system schematically shown by Fig.\ref{Fig:Fig.1}
inspired in the experiments reported by Refs.\cite{Kouwenhoven2023,MinimalModel2}.
Distinctly, we account for the fine structure due to a quantum spin
$S$. The simplest way to introduce such is via the \textit{Ising-like}
Hamiltonian $JS^{z}s^{z}$\cite{Ising}. {To that
end, we adopt the exchange term $J$ between the}\textit{{{}
spin-z}}{{} operators $S^{z}=\sum_{m}m|m\rangle\langle m|,$
wherein $m=[-S,-S+1,...,S-1,S],$ and $s^{z}=\frac{1}{2}\sum_{\sigma}\sigma d_{L\sigma}^{\dagger}d_{L\sigma}$}\textit{{{}
}}{for the quantum spin $S$ and left QD, respectively,
with $d_{L\sigma}^{\dagger}(d_{L\sigma})$ as the creation (annihilation)
operator and $\sigma=\pm1(\uparrow,\downarrow)$. For the right QD,
we have $d_{R\sigma}^{\dagger}(d_{R\sigma}).$}

The QDs are found within the spinless regime, where we arbitrary choose
the spin-up channel $\sigma=\uparrow$ as relevant, due to an imposed
large Zeeman splitting. Such a dimer of QDs then constitutes the minimal
Kitaev chain, being the QDs linked to each other by means of ECT and
CAR mechanisms, described by the hopping $t$ and superconducting
pairing $\Delta$ terms, respectively. This scenario is mimicked by
the effective Hamiltonian
\begin{eqnarray}
{\cal {H}} & = & (\mu_{L\uparrow}+\frac{J}{2}S^{z})d_{L\uparrow}^{\dagger}d_{L\uparrow}+\mu_{R\uparrow}d_{R\uparrow}^{\dagger}d_{R\uparrow}+(td_{L\uparrow}d_{R\uparrow}^{\dagger}\nonumber \\
 & + & \Delta d_{L\uparrow}d_{R\uparrow}+\text{{H.c.}}),\label{eq:H1}
\end{eqnarray}
where $\mu_{L\uparrow(R\uparrow)}$ represents the chemical potential
for the QD $\alpha=L,R.$ Both the electronic operators of the chain
can be expressed in the\textit{ }MBS basis $\gamma_{L1(L2)}$ and
$\gamma_{R1(R2)}$ for the left and the right QDs, respectively. To
perform such, it is imperative to evoke the relations $d_{L\uparrow}=(\gamma_{L1}+i\gamma_{L2})/\sqrt{2}$
and $d_{R\uparrow}=(\gamma_{R1}+i\gamma_{R2})/\sqrt{2},$ which yield
\begin{align}
{\cal {H}} & =(\mu_{L\uparrow}+\frac{J}{2}S^{z})(\frac{1}{2}+i\gamma_{L1}\gamma_{L2})+\mu_{R\uparrow}(\frac{1}{2}+i\gamma_{R1}\gamma_{R2})\nonumber \\
 & +i(\Delta-t)\gamma_{L1}\gamma_{R2}+i(\Delta+t)\gamma_{L2}\gamma_{R1},\label{eq:MajoranaChain}
\end{align}
{from where the }\textit{{``Majorana
chain regime''}}{{} $t=\Delta$ emerges, being characterized
by a }\textit{{tight-binding-like}}{{}
Hamiltonian comprising the }\textit{{Majoranas}}{{}
of the system following the chain sequence $\gamma_{L1},\gamma_{L2}$
and $\gamma_{R1}$ (MF trimer) or $\gamma_{L2}$ and $\gamma_{R1}$
(MF dimer). In the }\textit{{sweet spot}}{\cite{Flensberg_2012(Poor)},
a limiting case within the }\textit{{``Majorana chain
regime''}}{{} with the constraints $\mu_{\alpha\uparrow}=\mu_{\bar{\alpha}\uparrow}=J=0$,
wherein $\bar{\alpha}=L(R)$ corresponds to the opposite QD $\alpha=R(L),$
spatially isolated PMMs $\gamma_{L1}$ and $\gamma_{R2}$ appear at
the left and right QDs, respectively\cite{Flensberg_2012(Poor),Kouwenhoven2023,MinimalModel2}.}

{We call particular attention to the exchange term
$J$ that plays the role of an effective chemical potential on the
left QD. It means that beyond the gate voltage variations attached
to the QD triggering the PMM spillover into the opposite QD, similar
behavior is expected to occur due to $J.$ However, as the quantum
spin $S$ can be fermionic or bosonic, we shall see peculiar transition
spectra depending on the spin statistics and interplay with the superconducting
pairing $\Delta.$ }

{In which concerns quantum computing aspects of PMMs,
in Ref.~\cite{Poor3} a protocol to initialize a PMM-based qubit state
via the detuning of both the energy levels of the QDs is deeply discussed
and can be, in principle, extended to our system. As the exchange
$J$ allows the tuning of the left QD energy with intricate spin statistics
dependence, a second exchange term, let us say $J_{R},$ from an extra
spin coupled to the right QD, not solely would fulfill the QDs detuning
requirement, but would bring novel possibilities due to the }\textit{{Ising-like}}{{}
Hamiltonians of the system. However, we let the exploration of adding a
new spin to the system to be done elsewhere. Here, we focus on theoretical
spectral analysis and tunneling spectroscopic tool, revealing that
Kitaev dimers with PMM spillover due to an exchange coupling, can
be employed as a detector of the spin statistics from a quantum spin.}

\subsection{Spectral Analysis}

In order to describe the spillover of PMMs, the evaluation of energy
dependent retarded Green's functions (GFs) for the QDs $\alpha$ are
timely. To this end, we should consider the ordinary spectral densities ${\cal {A}}_{d_{\alpha\uparrow}d_{\alpha\uparrow}^{\dagger}}(\omega)=(-1/\pi)\text{Im}\langle\langle d_{\alpha\uparrow};d_{\alpha\uparrow}^{\dagger}\rangle\rangle$
and ${\cal {A}}_{d_{\alpha\uparrow}^{\dagger}d_{\alpha\uparrow}}(\omega)=(-1/\pi)\text{Im}\langle\langle d_{\alpha\uparrow}^{\dagger};d_{\alpha\uparrow}\rangle\rangle$,
where $\langle\langle A;B\rangle\rangle$ stands for the corresponding
GF. The anomalous GFs ${\cal {A}}_{d_{\alpha\uparrow}^{\dagger}d_{\alpha\uparrow}^{\dagger}}(\omega)=(-1/\pi)\text{Im}\langle\langle d_{\alpha\uparrow}^{\dagger};d_{\alpha\uparrow}^{\dagger}\rangle\rangle$
and ${\cal {A}}_{d_{\alpha\uparrow}d_{\alpha\uparrow}}(\omega)=(-1/\pi)\text{Im}\langle\langle d_{\alpha\uparrow};d_{\alpha\uparrow}\rangle\rangle$
should be taken into account too. As shown below, they determine the
MBS component ${\cal {A}}_{\gamma_{\alpha j}}(\omega)=(-1/\pi)\text{Im}\langle\langle\gamma_{\alpha j};\gamma_{\alpha j}\rangle\rangle$
of the QD. From $\gamma_{L1(L2)}$ and $\gamma_{R1(R2)},$ we obtain
the GF as follows
\begin{eqnarray}
\langle\langle\gamma_{\alpha j};\gamma_{\alpha j}\rangle\rangle & = & \frac{1}{2}[\langle\langle d_{\alpha\uparrow};d_{\alpha\uparrow}^{\dagger}\rangle\rangle+\langle\langle d_{\alpha\uparrow}^{\dagger};d_{\alpha\uparrow}\rangle\rangle\nonumber \\
 & + & \epsilon_{j}(\langle\langle d_{\alpha\uparrow}^{\dagger};d_{\alpha\uparrow}^{\dagger}\rangle\rangle+\langle\langle d_{\alpha\uparrow};d_{\alpha\uparrow}\rangle\rangle)],\label{eq:Expansion}
\end{eqnarray}
where $\epsilon_{j}=+1,-1$ for $j=1,2.$ Thus, the task of calculating
the GFs of Eq.(\ref{eq:Expansion}) can be achieved via the standard
equation-of-motion (EOM) approach\cite{Flensberg-book}, which is
summarized as
\begin{eqnarray}
(\omega+i\Gamma)\langle\langle A;B\rangle\rangle=\langle[A,B]_{+}\rangle+\langle\langle[A,{\cal {\cal {H}}}];B\rangle\rangle,\label{eq:EOM}
\end{eqnarray}
where $\Gamma$ mimics the natural broadening, supposed to be symmetric
for simplicity, arising from the outside environment. By applying
the EOM technique to Eq.(\ref{eq:H1}) for $t\neq\Delta$ we then
find
\begin{eqnarray}
\langle\langle d_{\alpha\uparrow};d_{\alpha\uparrow}^{\dagger}\rangle\rangle & = & \frac{1}{2S+1}\sum_{m}\frac{1}{\omega+i\Gamma-\mu_{\alpha\uparrow}-\frac{Jm}{2}\delta_{\alpha L}-\Sigma_{\alpha}^{+}},\nonumber \\
\label{eq:GF1}
\end{eqnarray}
\begin{eqnarray}
\langle\langle d_{\alpha\uparrow}^{\dagger};d_{\alpha\uparrow}\rangle\rangle & = & \frac{1}{2S+1}\sum_{m}\frac{1}{\omega+i\Gamma+\mu_{\alpha\uparrow}+\frac{Jm}{2}\delta_{\alpha L}-\Sigma_{\alpha}^{-}},\nonumber \\
\label{eq:GF2}
\end{eqnarray}
\begin{eqnarray}
\langle\langle d_{\alpha\uparrow}^{\dagger};d_{\alpha\uparrow}^{\dagger}\rangle\rangle & = & \frac{\eta_{\alpha}}{2S+1}\sum_{m}\frac{2t\Delta K_{\alpha}^{-}}{\omega+i\Gamma+\mu_{\alpha\uparrow}+\frac{Jm}{2}\delta_{\alpha L}-\Sigma_{\alpha}^{-}},\nonumber \\
\label{eq:GF3}
\end{eqnarray}
and
\begin{eqnarray}
\langle\langle d_{\alpha\uparrow};d_{\alpha\uparrow}\rangle\rangle & = & \frac{\eta_{\alpha}}{2S+1}\sum_{m}\frac{2t\Delta K_{\alpha}^{+}}{\omega+i\Gamma-\mu_{\alpha\uparrow}-\frac{Jm}{2}\delta_{\alpha L}-\Sigma_{\alpha}^{+}},\nonumber \\
\label{eq:GF4}
\end{eqnarray}
where we used $\langle\langle A;B\rangle\rangle=\sum_{m}\langle\langle A\left|m\right\rangle \left\langle m\right|;B\rangle\rangle,$
the thermal average $\left\langle \left|m\right\rangle \left\langle m\right|\right\rangle =\frac{1}{2S+1}$,
$\delta_{\alpha L}$ as the Kronecker Delta and $\eta_{\alpha}=-1,+1$
for $\alpha=L,R,$ respectively. The self-energy correction due to
the couplings $t,\Delta$ and $J$ is $\Sigma_{\alpha}^{\pm}=\tilde{K}_{\bar{\alpha}}^{\pm}+(2t\Delta)^{2}K_{\bar{\alpha}}K_{\alpha}^{\pm},$
with
\begin{eqnarray}
\tilde{K}_{\alpha}^{\pm}=\frac{(\omega+i\Gamma)(t^{2}+\Delta^{2})\pm(\mu_{\alpha\uparrow}+\frac{Jm}{2}\delta_{\alpha L})(t^{2}-\Delta^{2})}{(\omega+i\Gamma)^{2}-(\mu_{\alpha\uparrow}+\frac{Jm}{2}\delta_{\alpha L})^{2}},\nonumber \\
\label{eq:SE1}
\end{eqnarray}
\begin{equation}
K_{\alpha}=\frac{\omega+i\Gamma}{(\omega+i\Gamma)^{2}-(\mu_{\alpha\uparrow}+\frac{Jm}{2}\delta_{\alpha L})^{2}}\label{eq:SE2}
\end{equation}
and
\begin{equation}
K_{\alpha}^{\pm}=\frac{K_{\bar{\alpha}}}{\omega+i\Gamma\pm\mu_{\alpha\uparrow}\pm\frac{Jm}{2}\delta_{\alpha L}-\tilde{K}_{\bar{\alpha}}^{\mp}}.\label{eq:SE3}
\end{equation}

{It is worth mentioning that the expression of Eq.(\ref{eq:Expansion})
for the }\textit{{Majorana}}{{} GF $\langle\langle\gamma_{\alpha j};\gamma_{\alpha j}\rangle\rangle$
wrap ups processes as electron (hole) tunneling $\langle\langle d_{\alpha\uparrow};d_{\alpha\uparrow}^{\dagger}\rangle\rangle$
($\langle\langle d_{\alpha\uparrow}^{\dagger};d_{\alpha\uparrow}\rangle\rangle$)
and the corresponding local Andreev reflection $\langle\langle d_{\alpha\uparrow}^{\dagger};d_{\alpha\uparrow}^{\dagger}\rangle\rangle$
($\langle\langle d_{\alpha\uparrow};d_{\alpha\uparrow}\rangle\rangle$)
at the QD $\alpha.$ As the coefficient $\epsilon_{j}$ changes sign
when the }\textit{{Majorana}}{{} index
$j$ is swapped, the aforementioned transport channels can interfere
constructively or destructively depending on the }\textit{{Majorana}}{{}
GF for a fixed QD. In this way, the spectral function ${\cal {A}}_{\gamma_{\alpha j}}(\omega)$
is expected to exhibit resonant peaks and dips spanned by $\omega$
and $J$ for tuned parameters $t,\Delta$ and $\mu_{\alpha\uparrow}.$
Interestingly enough, such interference processes lead to different
scenarios for the system }\textit{{``Majorana chain
regime'',}}{{} such as the MF trimer ($\gamma_{L1},\gamma_{L2}$
and $\gamma_{R1}$) and dimer ($\gamma_{L2}$ and $\gamma_{R1}$),
as well as the PMMs ($\gamma_{L1}$ and $\gamma_{R2}$) in the }\textit{{sweet
spot}}{. Therefore, the so-called bonding, anti-bonding
and non-bonding molecular states, and the zero-mode are due to the
quasiparticle interference encoded by ${\cal {A}}_{\gamma_{\alpha j}}(\omega).$
We shall see these features in Sec. 3.}

\subsection{Tunneling Spectroscopy}

{To access experimentally the features of the spectral densities,
one should perform a quantum transport evaluation of the system conductance.
From a theoretical perspective, we adopt the proposal found in Ref.~\cite{Conductance}, which consists in flanking
a QD by source and drain metallic leads with chemical potentials $\mu_{\text{{Source}}}$
and $\mu_{\text{{Drain}}},$ respectively. Thus, we should add to
the Hamiltonian of Eq.(\ref{eq:H1}) the terms for the leads and their
couplings to the left QD as follows}

{
\begin{equation}
{\cal {H}}_{\text{{Total}}}={\cal {H}}+\sum_{\tilde{q},\textbf{k}}(\varepsilon_{\textbf{k}}-\mu_{\tilde{q}})c_{\tilde{q}\textbf{k}}^{\dagger}c_{\tilde{q}\textbf{k}}+V_{L}\sum_{\tilde{q},\textbf{k}}(c_{\tilde{q}\textbf{k}}^{\dagger}d_{L\uparrow}+\text{{H.c.}}),\label{eq:HTotal}
\end{equation}
where the second term stands for the leads with $\tilde{q}=q=\text{{Source}}$
($\tilde{q}=\bar{q}=\text{{Drain}}$) and in the third part $V_{L}$
represents the symmetric QD-lead tunneling term. In this way, the current
$I_{q}$ can be decomposed into the parts
\begin{equation}
I_{q}  = I_{q}^{\text{ET}}+I_{q}^{\text{LAR}}+I_{q}^{\text{CAR}},\label{eq:Current1}
\end{equation}
where }

{
\begin{equation}
I_{q}^{\text{ET}} = \frac{e}{h}\int d\varepsilon\tau_{q\bar{q}}^{\text{ET}}(\varepsilon)\left[f_{q}^{e}(\varepsilon)-f_{\bar{q}}^{e}(\varepsilon)\right],\label{eq:CurrentET}
\end{equation}
\begin{equation}
I_{q}^{\text{CAR}} = \frac{e}{h}\int d\varepsilon\tau_{q\bar{q}}^{\text{CAR}}(\varepsilon)\left[f_{q}^{e}(\varepsilon)-f_{\bar{q}}^{h}(\varepsilon)\right]\label{eq:CurrentCAR}
\end{equation}
and
\begin{equation}
I_{q}^{\text{LAR}}  =  \frac{e}{h}\int d\varepsilon\tau_{qq}^{\text{LAR}}(\varepsilon)\left[f_{q}^{e}(\varepsilon)-f_{q}^{h}(\varepsilon)\right],\label{eq:CurrentLAR}
\end{equation}
where $I_{q}^{\text{ET}}$ and $I_{q}^{\text{CAR}}$ stem from the
currents for the electron tunneling (ET) and crossed Andreev reflection
(CAR) with occupation probabilities of an electron $f_{\bar{q}}^{e}(\varepsilon)$
and hole $f_{\bar{q}}^{h}(\varepsilon)$ states at lead $\bar{q},$
respectively, where $f_{q}^{j}(\varepsilon)$ stands for the Fermi
distribution at lead $q$ and $j=e(h)$ for the electron (hole)
quasiparticle. In the case of the local Andreev reflection (LAR) $I_{q}^{\text{LAR}},$
the hole emission occurs into the same terminal $q,$ as it depends
on $f_{q}^{h}(\varepsilon).$ Particularly, with the assumption $\mu_{\text{{Source}}}=-\mu_{\text{{Drain}}}=\text{{eV}}/2$
we conclude that $f_{q}^{e}(\varepsilon)=f_{\bar{q}}^{h}(\varepsilon),$
which from Eq.(\ref{eq:CurrentCAR}) results in $I_{q}^{\text{CAR}}=0$
(this not implies in the absence of the system intrinsic CAR given
by $\Delta$ between left and right QDs) and $I_{q}=-I_{\bar{q}},$
whose the transmittance coefficients are determined by $\tau_{q\bar{q}}^{\text{{ET}}}=(2S+1)\Gamma_{L}^{2}|\langle\langle d_{L\uparrow};d_{L\uparrow}^{\dagger}\rangle\rangle_{\omega}|^{2}$
and $\tau_{qq}^{\text{{LAR}}}=(2S+1)\Gamma_{L}^{2}|\langle\langle d_{L\uparrow}^{\dagger};d_{L\uparrow}^{\dagger}\rangle\rangle_{\omega}|^{2},$
where $\Gamma_{L}=2\pi V_{L}^{2}\rho$ represents the symmetric electron
(hole)- QD spectral broadening term and $\rho$ the lead DOS. Thus,
the conductance from the source reservoir reads}

{
\begin{equation}
\mathcal{G}_{q}=\frac{dI_{q}}{d\text{{V}}}=\frac{dI_{q}^{\text{ET}}}{d\text{{V}}}+\frac{dI_{q}^{\text{LAR}}}{d\text{{V}}},\label{eq:Gtotal2}
\end{equation}
with }

{
\begin{eqnarray}
\frac{dI_{q}^{\text{ET}}}{d\text{{V}}} & = & \frac{e^{2}}{2h}\frac{1}{T}\int d\varepsilon\tau_{q\bar{q}}^{\text{ET}}(\varepsilon)\{f_{q}^{e}(\varepsilon)[1-f_{q}^{e}(\varepsilon)]\nonumber \\
 & + & f_{\bar{q}}^{e}(\varepsilon)[1-f_{\bar{q}}^{e}(\varepsilon)]\},
\end{eqnarray}
and
\begin{eqnarray}
\frac{dI_{q}^{\text{LAR}}}{d\text{{V}}} & = & \frac{e^{2}}{2h}\frac{1}{T}\int d\varepsilon\tau_{qq}^{\text{LAR}}(\varepsilon)\{f_{q}^{e}(\varepsilon)[1-f_{q}^{e}(\varepsilon)]\nonumber \\
 & + & f_{\bar{q}}^{e}(\varepsilon)[1-f_{\bar{q}}^{e}(\varepsilon)]\},
\end{eqnarray}
where we employed the identity $f_{q}^{e}(\varepsilon)=f_{\bar{q}}^{h}(\varepsilon)$ and }

{
\begin{equation}
\frac{\partial f_{q(\bar{q})}^{e}(\varepsilon)}{\partial\text{{V}}}=\pm\frac{e}{2T}f_{q(\bar{q})}^{e}(\varepsilon)[1-f_{q(\bar{q})}^{e}(\varepsilon)],\label{eq:Derivative}
\end{equation}
with $k_{B}=1,$ $f_{q}^{e}(\varepsilon)=f(\varepsilon-\text{{eV}}/2)$,
$f_{\bar{q}}^{e}(\varepsilon)=f(\varepsilon+\text{{eV}}/2)$ and $f(x)=1/(1+e^{x/T})$. }

{As $\frac{1}{T}f_{q(\bar{q})}^{e}(\varepsilon)[1-f_{q(\bar{q})}^{e}(\varepsilon)])=\left(-\frac{\partial f_{q(\bar{q})}^{e}(\varepsilon)}{\partial\varepsilon}\right)\rightarrow\delta(\varepsilon\mp\text{{eV}}/2)$
when $T\rightarrow0\text{{K},}$ then we find}

{
\begin{equation}
\mathcal{G}_{\text{{Source}}}=\mathcal{G}_{\text{{ET}}}(\text{{eV}})+\mathcal{G}_{\text{{LAR}}}(\text{{eV}}),\label{eq:GSource}
\end{equation}
}

{
\begin{equation}
\mathcal{G}_{\text{{ET}}}(\text{{eV}})=\frac{e^{2}}{2h}[\tau_{q\bar{q}}^{\text{{ET}}}(\text{{eV}}/2)+\tau_{q\bar{q}}^{\text{{ET}}}(-\text{{eV}}/2)],\label{eq:ET}
\end{equation}
and
\begin{equation}
\mathcal{G}_{\text{{LAR}}}(\text{{eV}})=\frac{e^{2}}{2h}[\tau_{qq}^{\text{{LAR}}}(\text{{eV}}/2)+\tau_{qq}^{\text{{LAR}}}(-\text{{eV}}/2)]\label{eq:LAR}
\end{equation}
in agreement with Ref.~\cite{Conductance}. Before the numerical analysis of
Sec. 3, we highlight that the GFs that determine $\mathcal{G}_{\text{{Source}}}$
follow the set of Eqs.(\ref{eq:GF1})-(\ref{eq:GF4}) with the substitution
$\Gamma\rightarrow\Gamma+\Gamma_{L}\delta_{\alpha L},$ due to the
coupling between the left QD and source-drain leads.}

\section{Results and Discussion}

\begin{figure}[!]
\centering\includegraphics[width=1\columnwidth]{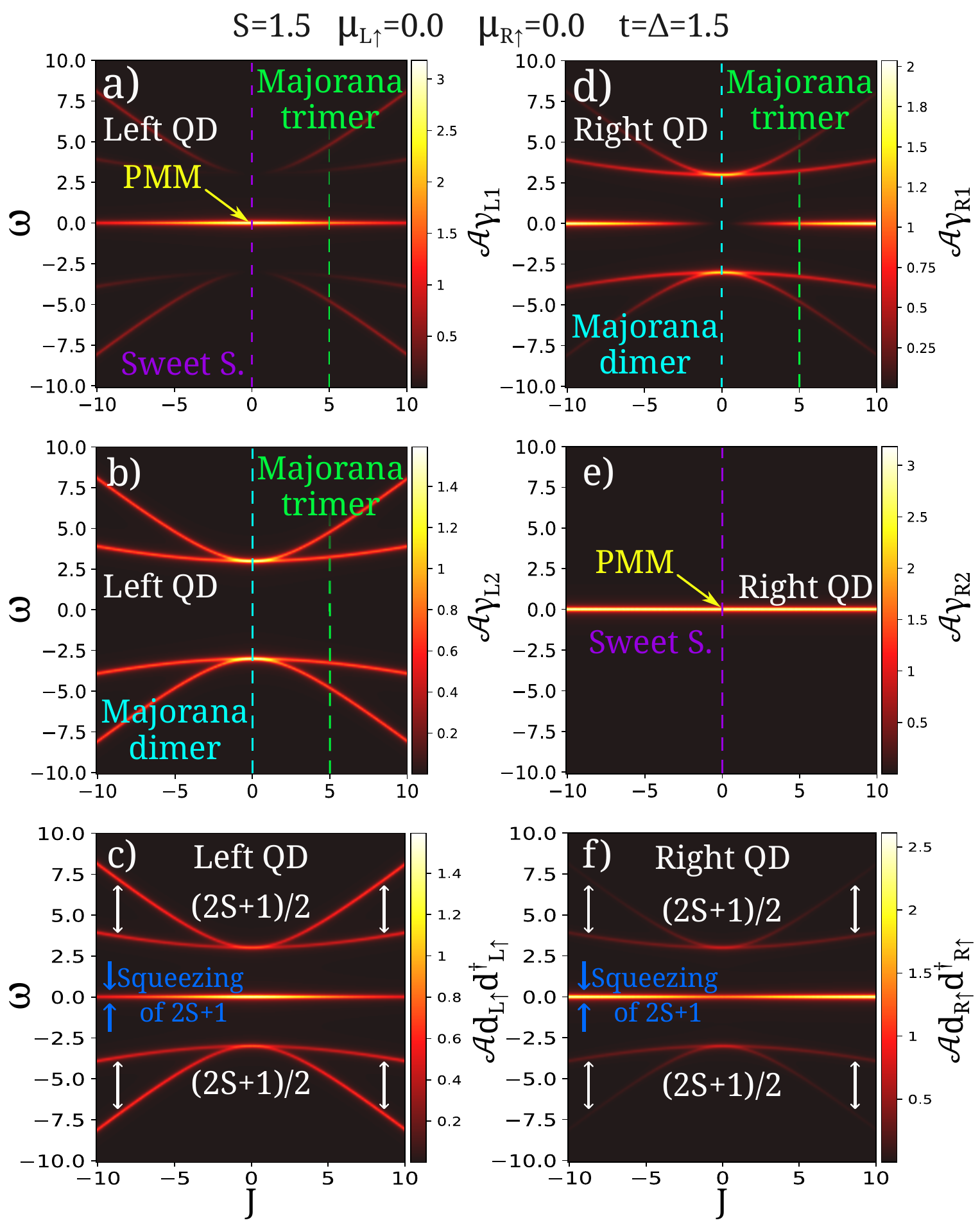} \caption{\label{Fig:Fig.2} The\textit{ }spillover of the PMM induced by $J.$
Color maps of ${\cal {A}}_{d_{\alpha\uparrow}d_{\alpha\uparrow}^{\dagger}}$
and ${\cal {A}}_{\gamma_{\alpha j}}$ in the \textit{\textquotedblleft Majorana
chain regime\textquotedblright} spanned by $\omega$ and $J$ showing
that in the presence of the fermionic spin $S$, $2S+1$ of the fine
structure is the explicit part. The other half $2S+1$ squeezes itself
at $\omega=0$ forming the delocalized MF zero-mode due to the PMM
spillover. }
\end{figure}

In Fig.\ref{Fig:Fig.2}, we introduce the PMM spillover induced by
$J.$ Experimentally speaking, \ensuremath{J} can be tuned by changing
the distance between the spin $S$ site and the leftmost QD. For such
an analysis, we first consider{{} }the \textit{``Majorana
chain regime}'' $t=\Delta=1.5,$ with $\mu_{L\uparrow}=\mu_{R\uparrow}=0$
in arbitrary units for the fermionic case $S=1.5$, and evaluate the
spectral functions spanned by $\omega$ and $J.$ Particularly, when
$J=0$ the \textit{sweet spot }is restored and we have the MF dimer
$\gamma_{L2}$ and $\gamma_{R1}$ described by ${\cal {H}}=2it\gamma_{L2}\gamma_{R1},$
and the isolated PMMs $\gamma_{L1}$ and $\gamma_{R2},$ once they
do not enter into ${\cal {H}},$ at the left and right QDs, respectively.
However for $J\neq0,$ the Hamiltonian turns into ${\cal {H}}\sim i\frac{J}{2}S^{z}\gamma_{L1}\gamma_{L2}+2it\gamma_{L2}\gamma_{R1}$
yielding the MF trimer established by $\gamma_{L1},\gamma_{L2}$ and
$\gamma_{R1},$ which drives the system into the regime where $\gamma_{L1}$
``spills over'' towards $\gamma_{R1}.$

To perceive such a scenario, we begin the analysis with ${\cal {A}}_{\gamma_{L1}}$
and ${\cal {A}}_{\gamma_{R2}}$ in Figs.\ref{Fig:Fig.2}(a) and (e),
which due to the $J=0$ condition, gives rise to the non-local PMMs
$\gamma_{L1}$ and $\gamma_{R2}$ (purple line cuts), respectively.
These PMMs are represented by the finite and equal amplitudes for
the zero-energy mode at $\omega=0.$ At $J=0,$ ${\cal {A}}_{\gamma_{L2}}$
and ${\cal {A}}_{\gamma_{R1}}$ display a split-peak structure as
a function of $\omega,$ once $\gamma_{L2}$ and $\gamma_{R1}$ build
the MF dimer (cyan line cuts in Figs.\ref{Fig:Fig.2}(b) and (d)).
For $J\neq0$ and $\omega\neq0$, extra bottom (bonding) and top (anti-bonding)
arcs rise in Figs.\ref{Fig:Fig.2}(a), (b) and (d) for ${\cal {A}}_{\gamma_{L1}},{\cal {A}}_{\gamma_{L2}},$
and ${\cal {A}}_{\gamma_{R1}},$ respectively, as a consequence of
the MF trimer formation (green line cuts). In Figs.\ref{Fig:Fig.2}(a)
and (d) for $J\neq0,$ the resonance at $\omega=0$ corresponds to
the non-bonding state of the MF trimer. Particularly, the multiplicity
of arcs is $2S+1$ due to the fermionic spin and it corresponds to
half of the total fine structure $2\times(2S+1)$ as we shall see
in the analysis off the\textit{ ``Majorana chain regime''}. The
spillover of $\gamma_{L1}$ on $\gamma_{R1}$ can be noted in Figs.\ref{Fig:Fig.2}(a)
and (d), where the unbalance ${\cal {A}}_{\gamma_{L1}}(0)<{\cal {A}}_{\gamma_{R1}}(0)$
is observed away from $J=0.$ In this manner, we reveal that the zero
mode of the PMM $\gamma_{L1}$ ``spills over'' from its QD into
the opposite. Additionally, it means that the zero-mode from a delocalized
MF at both the QDs emerges as aftermath of the squeezing of the other
half $2S+1.$

Thus, the unbalance also in ${\cal {A}}_{d_{\alpha\uparrow}d_{\alpha\uparrow}^{\dagger}}(0)<{\cal {A}}_{d_{\bar{\alpha}\uparrow}d_{\bar{\alpha}\uparrow}^{\dagger}}(0)$
off $J=0$ and the explicit $2S+1$ fine structure of Figs.\ref{Fig:Fig.2}(c)
and (f) consist the hallmarks of the spillover of a PMM in the presence of
a fermionic spin.

\begin{figure}[!]
\centering\includegraphics[width=1\columnwidth]{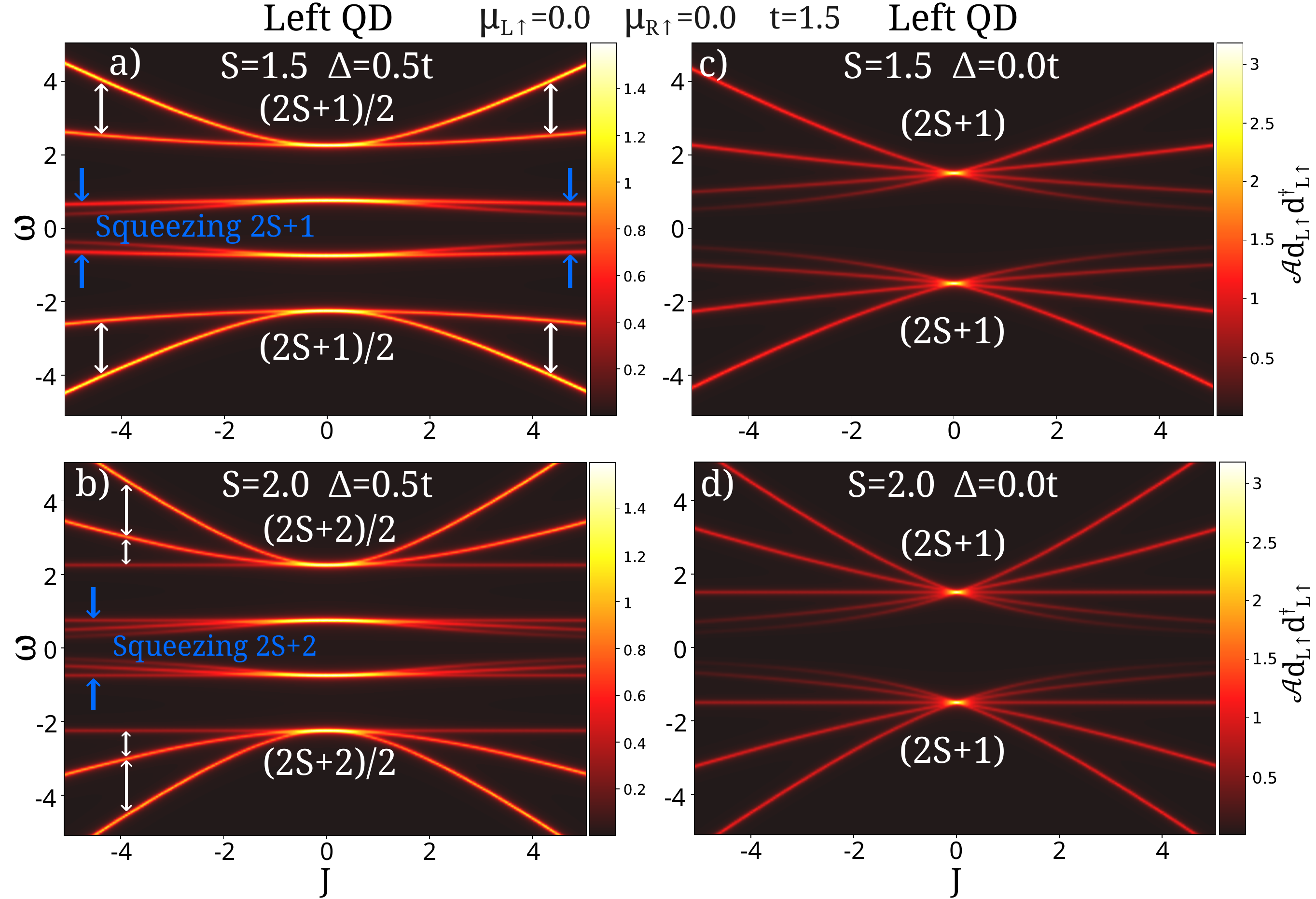} \caption{\label{Fig:Fig.3} Color map of ${\cal {A}}_{d_{L\uparrow}d_{L\uparrow}^{\dagger}}$
off the \textit{\textquotedblleft Majorana chain regime\textquotedblright}
spanned by $\omega$ and $J$ for: (a) Fermion case $S=1.5$. The
total fine structure contains $2\times(2S+1)$ levels as $|J|$ increases,
but the inner half reveals a squeezing into a delocalized MF zero-mode
due to the spillover of the PMM. (b) Boson case $S=2.$ The entire
fine structure is $2\times(2S+2)$ as $|J|$ increases and the inner
half shows the same trend of (a). (c) and (d) exhibit the dimer without
CAR $(\Delta=0)$ and the fine structure $2\times(2S+1)$ is statistics
independent.}
\end{figure}

In order to elucidate the dependence of fine structure on $S$ and
the underlying squeezing mechanism to build the zero-mode of the delocalized
MF due to the PMM spillover, in Fig.\ref{Fig:Fig.3} we present ${\cal {A}}_{d_{L\uparrow}d_{L\uparrow}^{\dagger}}$
for the case $\Delta=0.5t,$which corresponds to a situation off the
\textit{``Majorana chain regime''}. It is worth mentioning that
the exhibition of ${\cal {A}}_{d_{R\uparrow}d_{R\uparrow}^{\dagger}}$
is redundant, once it shares the same features observed in ${\cal {A}}_{d_{L\uparrow}d_{L\uparrow}^{\dagger}}.$

Particularly, the anti-crossing point at $J=0$ in Figs.\ref{Fig:Fig.3}(a)
and (b) of the energy levels arise from the existence of two MF dimers,
namely $\gamma_{L2},\gamma_{R1}$ and $\gamma_{L1},\gamma_{R2}$ as
aftermath of ${\cal {H}}=i(\Delta-t)\gamma_{L1}\gamma_{R2}+i(\Delta+t)\gamma_{L2}\gamma_{R1}.$
Thus, the Kitaev dimer with $\Delta\neq t$ and $\mu_{\alpha\uparrow}=\mu_{\bar{\alpha}\uparrow}=J=0$
exhibits a four-peak structure, which is distinct from the typical
two (bonding and anti-bonding) of an ordinary molecule with $\Delta=0,$
i.e., ${\cal {H}}=-it\gamma_{L1}\gamma_{R2}+it\gamma_{L2}\gamma_{R1}$
as can be noted in Figs.\ref{Fig:Fig.3}(c) and (d). Therefore, as
we can observe in Fig.\ref{Fig:Fig.3}(a) for the fermionic spin $S=1.5,$
$2\times(2S+1)$ levels appear as $|J|$ increases. This set reveals
a squeezing trend of the inner fine structure, which delimits precisely
$2S+1$ levels nearby $\omega=0,$ into a delocalized MF zero-mode.
We call attention to the pattern of this inner portion of fine structure
that precedes the \textit{``Majorana chain regime}'' $\Delta=t$
of Fig.\ref{Fig:Fig.2}, with all these inner levels squeezed at $\omega=0$
upon approaching $\Delta\rightarrow t.$

In Fig.\ref{Fig:Fig.3}(b) for the bosonic spin $S=2,$ the multiplicity
$2S+1$ changes to $2S+2.$ It means that the explicit part of the
fine structure becomes $2S+2$ when $t=\Delta$ and half of the total
spectrum $2\times(2S+2)$ squeezes at $\omega=0,$ i.e., $2S+2.$

To understand such a variation, let us remind that when the CAR is
turned-off $(\Delta=0),$ the fine structure $2\times(2S+1)$ takes
place and is independent of $S.$ This appears in Figs.\ref{Fig:Fig.3}(c)
and (d). As a matter of fact, the difference between fermions and
bosons for $\Delta\neq0$ relies on the mirror symmetry of the spectrum
around $\omega=|t|.$ In case of a bosonic $S,$ the multiplicity
$2S+1$ is odd and for $\Delta=0$ the state $m=0$ corresponds to
$\omega=|t|,$ being flanked by $S$ states. A finite $\Delta$ {[}Figs.\ref{Fig:Fig.3}(a)
and (b){]} then splits $|t|$ into $|\Delta-t|$ and $|\Delta+t|,$
thus changing $2S+1$ to $2S+2.$ As for fermionic $S,$ $2S+1$ is
even and the $m=0$ state does not exist, it cannot be split by $\Delta$
and the multiplicity $2S+1$ is maintained as aftermath. Such a behavior
can be viewed in the animated plots of the Supplementary Material.
Particularly for bosonic $S,$ notice that the frame $\Delta=0.01t$
defines the threshold for the peaks at $\omega=|t|$ (purple dashed
lines) to split into those at $|\Delta-t|$ (blue lines) and $|\Delta+t|$
(red lines), which can be seen for instance in the frame $\Delta=0.05t.$
In the other hand, there are no peaks at $\omega=|t|$ to be split
for fermionic $S.$

\begin{figure}[!]
\centering \includegraphics[width=1\columnwidth]{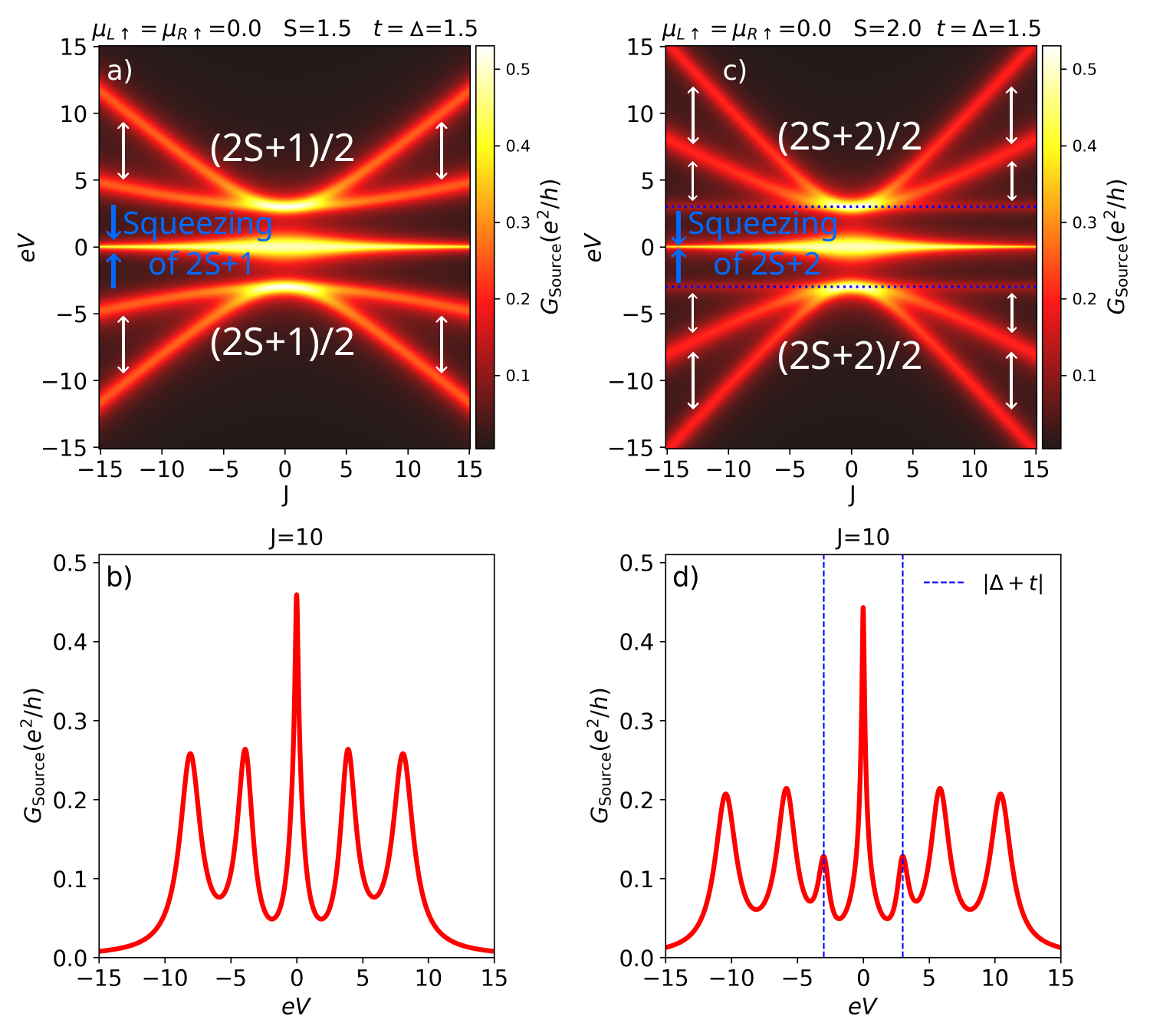} \caption{\label{Fig:Fig.4}{Differential conductance $\mathcal{G}_{\text{{Source}}}$
of Eq.(\ref{eq:GSource}) as a tunneling spectroscopic tool to detect
the spin statistics from a quantum spin with a Kitaev dimer and the
PMM spillover due to an exchange coupling $J.$ (a) Color map of $\mathcal{G}_{\text{{Source}}}$
spanned by the bias-voltage $\text{{eV}}$ and $J$ for the fermionic
case $S=1.5$ wherein energy levels bend (arcs) with $J$ and $(2S+1)/2$
side-peaks. (b) $\mathcal{G}_{\text{{Source}}}$ versus $\text{{eV}}$
for $J=10$ showing a zero-bias conductance $\mathcal{G}_{\text{{Source}}}(0)<e^{2}/2h,$
which strikes out a truly PMM, being a zero-energy Andreev state.
(c) Color map of $\mathcal{G}_{\text{{Source}}}$ for the bosonic
$S=2.0$ wherein low energy levels flanking the zero-bias peak do
not bend anymore with $J$ (flat dashed lines at $\text{{eV} }=|\Delta+t|$)
and $(2S+2)/2$ side-peaks. (d) $\mathcal{G}_{\text{{Source}}}$ versus
$\text{{eV}}$ for $J=10$ also showing a zero-energy Andreev state,
but with side-peaks at ${eV}=|\Delta+t|.$ The flat levels in (c)
represent the hallmark of a bosonic quantum spin.}}
\end{figure}

{Experimentally, one pathway to observe our results
consists in performing the tunneling spectroscopy approach of Sec. 2.2. Here, in the simulations, we adopt $\Gamma_{L}$ as energy unit
and $\Gamma=0.01$ in Eqs.(\ref{eq:GF1}) and (\ref{eq:GF3}) for
the left QD with $\Gamma\rightarrow\Gamma+\Gamma_{L}\delta_{\alpha L}$.
In this way, in Fig.\ref{Fig:Fig.4}(a) we depict the conductance $\mathcal{G}_{\text{{Source}}}$
of Eq.(\ref{eq:GSource}) spanned by the bias-voltage $\text{{eV}}$
and $J$ for the fermionic case $S=1.5,$ with $t=\Delta=1.5$ and
$\mu_{L\uparrow}=\mu_{R\uparrow}=0.$ Note that the conductance pattern
is qualitatively the same verified in Fig.\ref{Fig:Fig.2}(c) for
${\cal {A}}_{d_{L\uparrow}d_{L\uparrow}^{\dagger}}$ with explicit multiplicity
$2S+1.$ Particularly in Fig.\ref{Fig:Fig.4}(b), we
show the bias-voltage dependence of $\mathcal{G}_{\text{{Source}}}$
with $J=10.$ In this panel, we clearly observe $(2S+1)/2=2$ resonant
states for $S=1.5$ flanking the zero-bias peak, which according to
Figs.\ref{Fig:Fig.2} and \ref{Fig:Fig.3}, represent the net effect
of the squeezing of the other half of the fine structure as a zero-mode.
Additionally, as $\mathcal{G}_{\text{{Source}}}(0)$ stays below $e^{2}/2h$
we can safely conclude that the zero-energy state does not arise from
an isolated MF~\cite{Conductance}, which in this case would be a truly PMM.
Indeed, this zero-energy corresponds to a delocalized MF zero-mode
or simply an Andreev state due to the PMM spillover towards the right
QD. Noteworthy, to distinguish Figs.\ref{Fig:Fig.4}(a) and (b) from
a counterpart bosonic case, which has the same number of side-peaks
as in the fermionic situation $S=1.5$ but with the formula $(2S+2)/2=2$
for $S=1,$ we should inspect if the conductance profile contains
resonant states pinned at $|\Delta+t|$ regardless the strength of
the exchange $J.$ According to the revealed mechanism responsible
for the multiplicity change from $2S+1$ to $2S+2$ for a bosonic
quantum spin, the state $m=0,$ being centered at $\text{{eV}}=|t|,$
consists in the unique existent resonant state due to the odd parity
of the original fine structure $2S+1$ when the superconductivity
is absent. Then, the superconductor splits $\text{{eV}}=|t|$ into
$|\Delta-t|$ and $|\Delta+t|$ and consequently, it rises the multiplicity. When $t=\varDelta,$ solely
the peak positions at $|\Delta+t|=|2t|$ remain intact upon varying $J$, as we can see denoted
by dashed lines in Figs.\ref{Fig:Fig.4}(c) and (d). For the even parity of $2S+1$ from a fermionic quantum spin, the
states $\text{{eV}}=|t|$ are removed and the superconductor does
not have the state $m=0$ to be split.}

{In summary, if the color map of $\mathcal{G}_{\text{{Source}}}$
versus $\text{{eV}}$ and $J$ shows two flat levels on J flanking
the zero-bias peak such as the marked dashed lines in Fig.\ref{Fig:Fig.4}(c),
the transition spectrum belongs to a bosonic $S$ and we should apply
the formula ``$2S+2=$ total of peaks, except that from the zero-bias''
to figure out the value of $S.$ Otherwise, if these near-zero energy
levels bend with $J$ (a pair of arcs) in the conductance color map
as shown in Fig.\ref{Fig:Fig.4}(a), we then change in the left side
of the previous formula to $2S+1$ and find the corresponding fermionic
quantum spin instead.}

\section{Conclusions}

We demonstrate that in a system of two superconducting, spinless QDs
corresponding to the minimal Kitaev chain, the PMM exhibits the spillover
effect from one QD to the other when perturbed by an exchange coupling
with a quantum spin $S$. This effect results in a delocalized MF
zero-mode, comprising half of the fine structure induced by the exchange
interaction. The remaining fine structure, representing $2S+1$ or
$2S+2$ states, arises from the fermionic or bosonic nature of the
spin, respectively. For a dimer without CAR, this fine structure consistently
shows $2S+1$ levels, independent of the spin statistics. These findings
demonstrate the potential for analyzing minimal Kitaev chains through
quantum spin interactions, with implications for spin-related phenomena
and quantum computing.

\section{Acknowledgments}

We thank the Brazilian funding agencies CNPq (Grants. Nr. 302887/2020-2,
303772/2023-9, 311980/2021-0, and 308695/2021-6), the S\~ao Paulo Research
Foundation (FAPESP; Grant No. 2023/13467-6), Coordena\c{c}\~ao de Aperfei\c{c}oamento
de Pessoal de N\'ivel Superior - Brasil (CAPES) -- Finance Code 001
and FAPERJ process Nr. 210355/2018. LSR acknowledges the support from
the Icelandic Research Fund (Rann\'is), Grant No. 239552-051. LSR thanks
Unesp for their hospitality. HS acknowledges the project No. 2022/45/P/ST3/00467
co-funded by the Polish National Science Centre and the European Union
Framework Programme for Research and Innovation Horizon 2020 under
the Marie Sk\l odowska-Curie grant agreement No. 945339.


%

\end{document}